\def\year{2019}\relax
\documentclass[letterpaper]{article} 
\usepackage{aaai19}  
\usepackage{times}  
\usepackage{helvet}  
\usepackage{courier}  
\usepackage{url}  
\usepackage{graphicx}  
\usepackage{algorithm}
\usepackage{booktabs}
\usepackage{algorithmic}
\usepackage{comment}
\usepackage{wrapfig}
\usepackage[round]{natbib}
\usepackage{pgfplots}
\usepackage{subcaption}
\usepackage[T1]{fontenc}
\usepackage{tikz}
\usepackage{fancyhdr}
\usetikzlibrary{shapes,arrows}
\pgfplotsset{compat=1.14}

\tikzstyle{block} = [rectangle, draw, 
    text width=6em, text centered, minimum height=3em]
\tikzstyle{dashblock} = [rectangle, draw, 
    text width=6em, text centered,dashed, minimum height=3em,node distance=1.1cm]
\tikzstyle{line} = [draw, -latex]
\tikzstyle{cloud} = [draw, ellipse, node distance=2cm,
    minimum height=2em]

\nocopyright

\begin{document}
%
\title{Auditing News Curation Systems:\\ A Case Study Examining Algorithmic and Editorial Logic in Apple News}

\author{Jack Bandy\\
{Northwestern University}\\
jackbandy@u.northwestern.edu,
\And
Nicholas Diakopoulos\\
{Northwestern University}\\
nad@northwestern.edu}

\fancypagestyle{plain}{
\fancyhf{}
\renewcommand{\headrulewidth}{0pt} 
\chead{\small{\textit{To Appear in} Proceedings of the Fourteenth International AAAI Conference on Web and Social Media (ICWSM 2020)}}}

\maketitle\thispagestyle{plain}
\begin{abstract}
This work presents an audit study of Apple News as a sociotechnical news curation system that exercises gatekeeping power in the media. We examine the mechanisms behind Apple News as well as the content presented in the app, outlining the social, political, and economic implications of both aspects. We focus on the Trending Stories section, which is algorithmically curated, and the Top Stories section, which is human-curated. Results from a crowdsourced audit showed minimal content personalization in the Trending Stories section, and a sock-puppet audit showed no location-based content adaptation. Finally, we perform an extended two-month data collection to compare the human-curated Top Stories section with the algorithmically-curated Trending Stories section. Within these two sections, human curation outperformed algorithmic curation in several measures of source diversity, concentration, and evenness. Furthermore, algorithmic curation featured more ``soft news'' about celebrities and entertainment, while editorial curation featured more news about policy and international events. To our knowledge, this study provides the first data-backed characterization of Apple News in the United States. 
\end{abstract}

\section{Introduction}
Scholars have long recognized the broad implications of mediating the flow of information in society \citep{McCombs1972,Entman1993}. Moderation of content along this flow is often referred to as \textit{gatekeeping}, ``culling and crafting countless bits of information into the limited number of messages that reach people each day'' \citep{Shoemaker2009}. Gatekeeping power constitutes ``a major lever in the control of society'' \citep{Bagdikian1983}, especially since prominent topics in the news can become prominent topics among the general public \citep{McCombs2005}. 

The flow of information across society has become more complex through the digitization, algorithmic curation, and intermediation of the news \citep{Diakopoulos2019}. Sociotechnical content curation systems like Google, Facebook, YouTube, and Apple News play significant roles as gatekeepers; their algorithms, processes, and content policies directly influence the public's media exposure and intake. Considering this powerful influence on society, our work asks: \textit{How might we systematically characterize the gatekeeping function of algorithmic curation platforms?} To address this question, we develop a framework for auditing algorithmic curators, including aspects of the curation \textit{mechanism} (e.g. churn rate and adaptation) and the curated \textit{content} (e.g. outputted sources and topics). We apply this framework in an audit of Apple News. 

With more than 85 million monthly active users \citep{Feiner2019}, Apple News now has a significant (and growing) influence on news consumption. Our audit focuses on two of the most prominent sections of the app: the ``Trending Stories'' section, which is algorithmically curated, and the ``Top Stories'' section, which is curated by an editorial staff. We observe minimal evidence of localization or personalization within the Trending Stories section. We also find that, compared to Top Stories, Trending Stories exhibit higher source concentration and gravitate towards content related to celebrities, entertainment, and the arts -- what journalism scholars classify as ``soft news'' \citep{Reinemann2012}.

This paper presents (1) a conceptual framework for auditing content aggregators, (2) an application of that framework to the Apple News application, with tools and techniques for synchronizing crowdsourced data collection and using app simulation to get around a lack of APIs, and (3) results and analysis from our audit, which both corroborate and nuance the current understanding of how Apple News curates information. We discuss these findings and their implications for future audit studies of algorithmic news curators.

\section{Background \& Related Work\label{sec:related}}

\subsection{Algorithmic Accountability}
As algorithmic approaches replace and supplant human decisions across society, researchers have pointed out the importance of holding algorithms accountable \citep{Gillespie2014,Diakopoulos2015,Garfinkel2017}. One way to do this is the ``algorithm audit,'' which derives its name and approach from longstanding methods in the social sciences designed to detect discrimination \citep{Sandvig2014}. For example, algorithm audits have exposed discrimination in image search algorithms \citep{Kay}, Google auto-complete \citep{Baker2013}, dynamic pricing algorithms \citep{Chen2016}, automated facial analysis \citep{Buolamwini2018}, and word embeddings \citep{Caliskan2017}. Of particular relevance to our work are audit studies focusing on news intermediaries which aggregate, filter, and sort news content from primary publishers.

\subsection{Auditing Intermediaries}
Examples of algorithmic news intermediaries include social media websites (e.g. Facebook, Twitter, reddit), search engines (e.g. Google, Bing), and news aggregation websites (e.g. Google News). On each of these platforms, algorithms select and sort content for millions of users, thus wielding significant power as algorithmic gatekeepers. This content moderation has attracted critical attention \citep{Gillespie2018}, and spurred some researchers to audit intermediaries and check for discrimination, diversity, or filter-bubble effects.

\subsubsection{Social Media}
In the case of social media, \citet{Bakshy} investigated Facebook's News Feed, finding that user choices (e.g., click history, visit frequency) played a stronger role than News Feed's algorithmic ranking when it came to helping users encounter ideologically cross-cutting content. A 2009 study of Twitter showed that trending topics were more than 85\% news-oriented \citep{Kwak2010}. However, because of the high churn rate, trending topics can exhibit temporal coverage bias depending on when users visit the site \citep{Chakraborty2015}.

\subsubsection{Web Search}
As the most popular platform for online search, Google has been the subject of numerous audit studies, some of which reveal discriminatory results. For example, Google's autocomplete feature was shown to exhibit gender and racial biases \citep{Baker2013}, and its image search was shown to systematically underrepresent women \citep{Kay}. Other audits show that user-generated content such as Stack Overflow, Reddit, and particularly Wikipedia play a critical role in Google's ability to satisfy search queries \cite{Vincent2019}

Several studies have investigated potential political bias in Google's search results \citep{Robertson2018,Diakopoulos,Epstein2017}, however, further research is needed to understand the extent and causes of apparent biases. For example, Google's search algorithms may increase exposure to particular news sources (with left-leaning audiences) due to freshness, relevance, or greater overall abundance of content on the web \citep{Trielli}.

Some studies have also investigated Google for creating virtual echo chambers, which may affect democratic processes. While concern has mounted over the search engine's filter-bubble effects \citep{Pariser}, studies have thus far found limited supporting evidence for the phenomenon \citep{Puschmann2018,Flaxman,Hannak2013,Robertson2018}. An analysis of more than 50,000 online users in the United States showed that the search engine can actually ``increase an individual's exposure to material from his or her less preferred side of the political spectrum'' \citep{Flaxman}. Still, some search results may vary with respect to location \citep{Kliman-Silver2015}, an effect that has the potential to create geolocal filter bubbles. 

\subsubsection{Google News}
As early as 2005, researchers zeroed in on Google's news system to assess potential bias in the platform \citep{Ulken2005,Schroeder2005}, showing that even shortly after its introduction, scholars were troubled by the platform's potential effects on journalism and the public at large. For example, the Associated Press was concerned that Google used their content without providing compensation \citep{Gaither2005}, and early on, Google News was suspected to have a conservative bias \citep{Ulken2005}.

Google News still attracts critical attention more than a decade later, but concerns have shifted towards the possibility of filter bubbles (as was the case with Google's search engine). Two studies in particular have addressed such concerns: \citet{Haim2018} tested for personalization with manually-created user profiles, and \citet{Nechushtai2019} did so with real-world users. The former study discovered ``minor effects'' on content diversity from both explicit personalization (from user-stated preferences) and implicit personalization (using inferences from observed online behavior). \citet{Nechushtai2019} showed that users with different political leanings and locations ``were recommended very similar news,'' but their study presented a separate concern: just five news outlets accounted for 49\% of the 1,653 recommended news stories. This source concentration highlights the multifaceted implications of news aggregators, which we return to later.

\subsubsection{Apple News}
Apple News has begun to attract the interest of various stakeholders in industry and research. The New York Times wrote about the app in October 2018 \citep{Nicas2018}, focusing on Apple's ``radical approach'' of using humans to curate content instead of just algorithms. Despite its growth, monetization on the platform has thus far proven difficult and drawn criticism \citep{Davies2017,Dotan2018}. Slate reports that it takes 6 million page views in the Apple News app to generate the same advertising revenue as 50,000 page views on its website -- a more than hundredfold difference \citep{Oremus2018}.

A study of Apple News' editorial curation choices in June 2018 analyzed tweets and email newsletters from the editors, finding that larger publishers appeared far more often \citep{Brown2018}. In a followup study, screen recordings captured the Top Stories section in the United Kingdom to collect 1,031 total articles, 75\% of which came from just six publishers \citep{Brown2018a}. This paper builds on and adds to these studies in two important ways. First, we design and use a method for \textit{fully automated data collection}, rather than relying on manual coding of screen recordings. This method allows us to collect both Top Stories and Trending Stories in the United States, whereas \citep{Brown2018a} collected only Top Stories in the United Kingdom. Second, we examine mechanical aspects of Apple News in addition to examining content. By investigating mechanisms such as adaptation and update frequency, we reveal several intriguing design choices and curation patterns within the app.

\section{Apple News \label{sec:apple-news}}
In this section, we contextualize the audit with an overview of Apple News' origin, evolution, and functionality.

A massive potential readership, low entry costs, an editorial staff, and declining traffic from other platforms are a few factors that have made Apple News a prominent figure in the news ecosystem within the last four years. Apple released the app in September 2015, making it just a few taps away from more than 1 billion iOS devices. Apple reported more than 85 million monthly active users in early 2019 \citep{Feiner2019}. As the publishing environment shifts, referral traffic from Apple News to publishers has surged \citep{Oremus2018,Tran}. Facebook's modified News Feed algorithms, which promote friends over news outlets, left a void that media executives hope Apple News can fill \citep{Weiss}. Besides the large iOS user base and potential for traffic, Apple provides an accessible publishing workflow. Publishers can create their own channel(s) and release articles using either the Apple News Format, RSS, or their own content management system. However, while these are flexible options for basic integration with Apple News, the editors only features content if it is in Apple News Format \citep{AppleInc.}.




\begin{figure}[!t] 
    \begin{center}
        
        \includegraphics[width=0.22\textwidth]{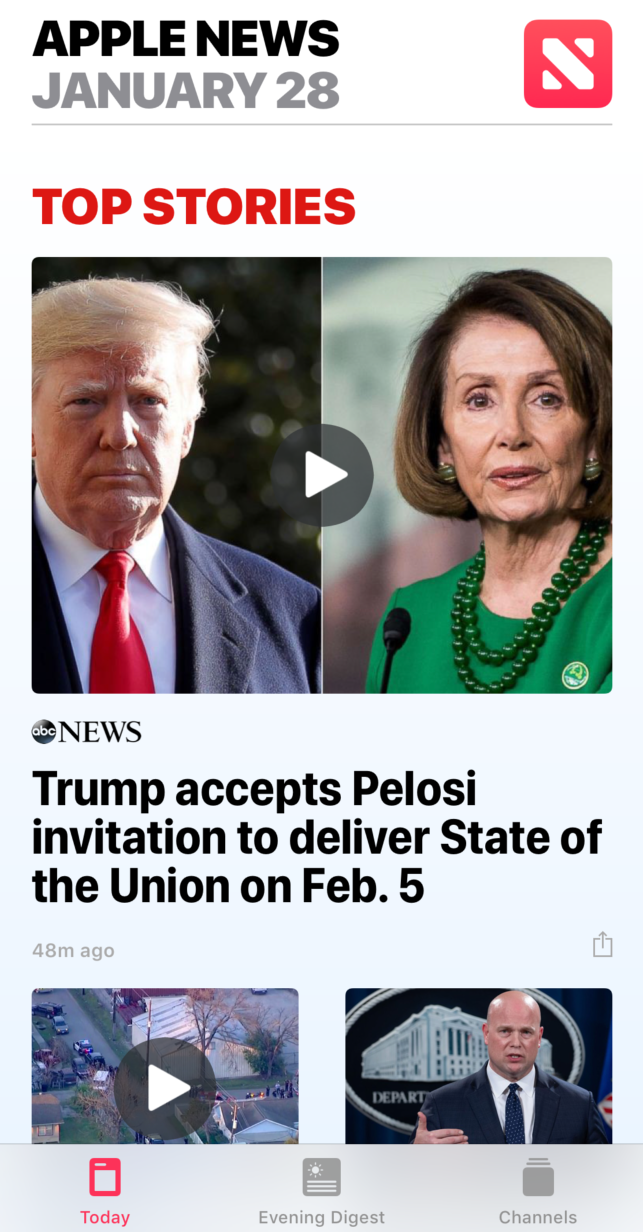}    		\includegraphics[width=0.22\textwidth]{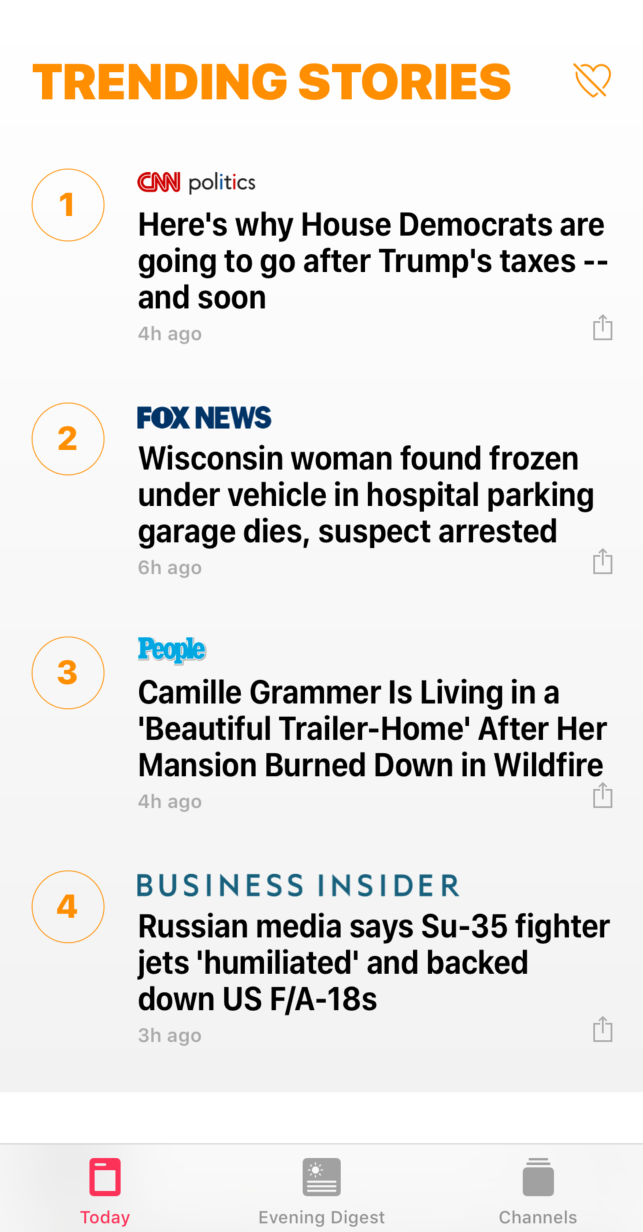}
    \end{center}
    \caption{Two screenshots from the Apple News app taken on an iPhone X simulator. The \textit{Top} Stories section is shown on the left, and the \textit{Trending} Stories section on the right.}
    \label{fig-screenshots}
\end{figure}

The Apple News app features a variety of content: top stories, popular stories, and stories relevant to a user's personal interests. Even though Apple has released several redesigns of the app, including one in March 2019, the various designs consistently make some sections more important and impactful. The primary tab (titled ``Today'') shows headlines and topical news from the current day. The second tab generally featured longer-form ``digest'' material before being renamed to ``News+'' in March 2019, when it was also redesigned to feature magazine content. A third tab (renamed from ``Channels'' to ``Following'' in March 2019) displays a list of publishers and news topics to view, like, or block. 

Notably, the primary tab highlights five aggregated ``Top Stories'' from Apple's editorial staff, which selects from about 100-200 pitches each day \citep{Nicas2018}. Thus, rather than a search engine or a news feed algorithm determining the prominence of a given story, the most prominent stories in Apple News are chosen by a staff of human editors. This editorial approach is a distinctive feature, which Apple CEO Tim Cook positioned as a way to combat the ``craziness'' of digital news, adding that human curation is not intended to be political, but instead to ensure the platform avoids ``content that strictly has the goal of enraging people'' \citep{Burch2018}. Still, while human editors curate the Top Stories section of the app, algorithms are used to determine what appears in the Trending Stories section \citep{Nicas2018}.

The Trending Stories section appears in the ``Today'' tab, below the Top Stories section (see Figure \ref{fig-screenshots}), and prominently displays a list of algorithmically-selected content. The first two Top Stories and the first two Trending Stories also appear in the Apple News widget, which is displayed by default when swiping left on the first iOS home screen.

Since at least 85 million users regularly see content in Top Stories and Trending Stories, we aim to better understand the mechanisms behind them and the content they display. By performing a focused audit on these two sections, we hope to characterize some of the ways Apple News is mediating human attention in today's news ecosystem.

\section{Audit Framework}\label{sec-framework}

In this section we develop and motivate a conceptual framework to guide our audit of Apple News, which we intend to capture important aspects of news curation systems. The framework is especially motivated by the way these systems can express editorial values and potentially influence personal, political, and/or economic dimensions of society through phenomena such as information overload, agenda-setting, and/or impacting  revenue for publishers.

It has become increasingly clear that algorithmic tools are ``heterogeneous and diffuse sociotechnical systems, rather than rigidly constrained and procedural formulas'' \citep{Seaver2017}. Therefore, our framework accounts for the role humans may play in such tools, and not merely their technical components. Contrary to the notion of ``neutral'' or ``purely technical,'' systems inevitably embed, embody, and propagate human values and biases \citep{Weizenbaum1976,Friedman1996,Introna2000,ONeil}, such as, in this case, editorial values about what and when to publish. As gatekeepers, news curation systems operate levers that  influence news distribution and consumption. These levers include the algorithms, content policies, and presentation design (i.e. layout) of the system.

Based on our review of prior studies and audits of platforms, we observe at least three important aspects of an intermediary such as Apple News that an audit should examine: (1) mechanism, (2) content, and (3) consumption. These three dimensions loosely align with the goals of better understanding (1) editorial bias arising as a result of a technological apparatus (i.e. mechanism), (2) exposure to information provided through a sociotechnical editorial process (i.e. content availability), and (3) attention patterns which encompass behavior of end users (i.e. consumption). Below is the framework along with some motivating questions:

\begin{itemize}
\item \textbf{Mechanism}
\begin{itemize}
\item How often does the curator update the content displayed? When do updates typically occur?
\item To what extent does the curator adapt content across users: is content personalized, localized, or uniform?
\end{itemize}

\item \textbf{Content}
\begin{itemize}
\item What sources does the curator display most often?
\item What topics does the curator display most often?
\end{itemize}

\item \textbf{Consumption}
\begin{itemize}
\item How does an appearance in the curator affect the attention given to a story?

\end{itemize}
\end{itemize}

\subsubsection{Mechanism}
While particularities of the mechanism may seem inconsequential, aspects such as update frequency and degree of adaptation play a crucial role in determining the type and quality of content a person sees in the curator. For example, the curator's update frequency can determine whether it emphasizes \textit{recent} content or \textit{relevant} content \citep{Chakraborty2015}, that is, information of present importance or longer-term importance. The churn rate for popularity-ranked content lists (such as Trending Stories) also effects the quality of that content, because high-quality content may not ``bubble up'' if the list is updated too frequently \citep{Salganik,Chaney2019,Ciampaglia2018}. Finally, more frequent updates increase the overall throughput of information, contributing to a phenomenon commonly known as ``information overload,'' whereby excessive amounts of content harms a person's ability to make sense of information \citep{Himma,Bawden2009}.

The political implications of news personalization mechanisms have also been pointed out in a wave of concern about the various threats to democratic efficacy posed by the ``filter bubble'' \citep{Pariser,Bozdag2015,Helberger2018,Flaxman}. These concerns range from diminished individual autonomy and undermined civic discourse to a lack of diverse exposure to information which may limit awareness and ability to contest ideas. While these concerns have recently been met with evidence that the filter bubble phenomenon is relatively minimal in practice \citep{Haim2018,Trilling2018}, evolving algorithmic curators need ongoing investigation to characterize exposure to differences of ideas or sources \citep{Diakopoulos2019}.

\subsubsection{Content}
In addition to mechanistic qualities such as update frequency and adaptation to different users, a curator's inclusion and exclusion of content is also of interest. The selection of news headlines represents ``a hierarchy of moral salience'' \citep{Schudson1995}. This is because many editorial decisions play a role in agenda-setting, sending topics from the mass media's attention into the general public's attention \citep{McCombs2005}. This relationship to the public discourse means that a curation system's content can have substantive political implications.

The content in a given curation system can be evaluated for source diversity, content diversity, and exposure diversity: variation in a system's information providers, topics/ideas/perspectives represented, and audience consumption behavior, respectively \citep{Napoli2011}. Investigating the diversity of news curators helps in understanding how they are impacting public discourse and democratic efficacy \citep{Nechushtai2019}. For instance, exposure diversity can help sustain democracy by supporting individual autonomy, creating a more inclusive public debate, and introducing new ways of contesting ideas \citep{Helberger2018}.

In addition to political implications, the sources of content in a curation system such as Apple News also has economic implications. For instance, publishing businesses have come to recognize the economic benefits of ranking well in Google search results, which drives more clicks, advertisement views, and therefore more revenue. Similarly, a story that ``ranks well'' in Apple News converts to revenue, at least to some extent, for the story's publisher. Since all 85 million users see the same content in several prominent sections of the app (e.g., Top Stories, Top Videos), an appearance in one of these sections can lead to a boost in traffic which can in turn increase revenue \citep{Nicas2018,Oremus2018,Dotan2018}.

\subsubsection{Consumption}
The potential for impact on user exposure to diverse content as well as on revenue for publishers makes it helpful to know about actual news \textit{consumption} resulting from a news curator, namely, how an appearance in the curator affects the attention given to a story. While monetizing in-app traffic has proven difficult thus far, publishers report substantial website traffic from Apple News referrals, which is already influencing revenue streams \citep{Nicas2018,Oremus2018,Tran,Dotan2018,Weiss}.  The current study, however, does not investigate such effects because Apple heavily restricts access to the relevant data, especially data about in-app behavior. Future work may consider alternative methodological approaches, such as panel studies, surveys, or data donations, to better understand the attentional effects of Apple News compared to other systems. For example, in an audit study of the Google Top Stories box, \cite{Trielli} were able to obtain timestamped referral data for 2,639 news articles and quantify the increase in referral traffic associated with appearances in different positions of the Top Stories box.

Having clarified how the mechanism, content, and consumption effects of news curation systems can have personal, political, and economic impact, we now apply this conceptual audit framework to Apple News. Again, because Apple closely guards the data needed to directly investigate specific effects on news consumption, our audit of Apple News focuses particularly on mechanism and content. 

\section{Audit Methods \label{sec:methods}}

In this section we detail the methods used to answer the questions from our conceptual audit framework, first addressing the primary tools used (Amazon Mechanical Turk and Appium), and then detailing our experiments. We will later discuss how future audit studies can and should employ these various methods. We investigate the mechanism behind the Trending Stories section first, because by knowing aspects of the algorithm such as its update frequency and its degree of personalization, we can better design and more efficiently execute the content audit phase. For example, if content is not personalized, then we can collect data from a single source instead of applying a crowdsourced audit. Once the mechanism is clarified, we design the content audit accordingly.

\subsection{Auditing Tools}

\subsubsection{Crowdsourced Auditing via Mechanical Turk}
For crowdsourced auditing, we leveraged Amazon Mechanical Turk (AMT), for which we designed a Human Intelligence Task (HIT) to collect screenshots of Apple News from workers on the platform. Our Institutional Review Board (IRB) reviewed the HIT description and determined it was not considered human research. Still, an important ethical consideration in using AMT is setting the wage for completing a task \citep{Hara2018}. To ensure we compensated workers according to the United States federal minimum wage of \$7.25 per hour, we deployed a pilot version of our HIT to fifteen users and measured the median time to completion, which was 6 minutes. We round up and consider \$0.75 to be the \textit{minimum} pay for the work of taking a  screenshot of the app for our data collection.

To audit the algorithm for personalization and eliminate time as a potential cause of variation in content, we needed multiple users to take screenshots in synchrony. Since crowdworkers perform tasks on their own schedule, synchronization represents a significant challenge. We therefore modify an approach from \citet{Bernstein2012}, in which we incentivize workers to wait until a designated time by offering a ``high-reward'' wage of \$4.00 per screenshot. For instance, users checking in at 10:36am and 10:42am would both be instructed to wait until 11:00am to take the screenshot in exchange for the increased wages. Once users uploaded their screenshot, we manually verify that the time displayed in the screenshot is the time we requested, then record the headlines shown.

After iterating on the task to ensure consistency in the data collection process, the task ultimately instructed workers to take the following steps: 

\begin{enumerate}
\item Unblock any previously blocked channels in Apple News
\item Force-close the Apple News app
\item Wait until the time was a round hour (e.g. 1:00, 2:00, etc.)
\item Re-open Apple News and take a screenshot showing the Trending Stories section
\item Upload the screenshot to Imgur and provide the URL to the screenshot, as well as input the zipcode of the device at the time the screenshot was taken

\end{enumerate}

\subsubsection{Sock Puppet Auditing via Appium}
Since the Apple News app lacks public APIs and implements security measures such as SSL pinning, we could not collect data via direct scraping. Crowdsourced auditing circumvents this and is ideal for auditing personalization, but to collect data for other experiments, we used Appium. Appium allows automatic control of the iPhone simulator on macOS, and was designed mainly for software verification. Appium's API can perform a number of actions in the iOS simulator, such as opening and closing applications, scrolling, finding and tapping buttons, and more. Using the API, we designed and built a suite of Python programs to collect stories from and run experiments in the Apple News app. We make our programs available with this paper\footnote{\url{https://github.com/comp-journalism/apple-news-scraper}}. The programs work by automatically opening the Apple News app, scrolling to Trending Stories, copying the ``share'' links to each of the stories, and saving them to a .csv file. Appium's API also allowed us to change the geolocation of the device, a feature we used in an experiment to test for localization. These experiments are detailed in the following section.

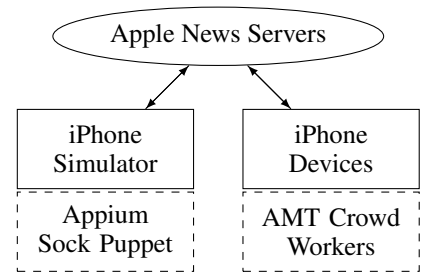
\begin{figure}[t]
    \centering
\begin{tikzpicture}[node distance = 2cm, auto]
    \node [cloud] (init) {Apple News Servers};
    \node [below of=init,node distance=1.5cm] (dummy) {};
    \node [block, left of=dummy,node distance=1.5cm] (sim) {iPhone Simulator};
    \node [block, right of=dummy,node distance=1.5cm] (dev) {iPhone Devices};
    \node [dashblock, below of=sim] (app) {Appium Sock Puppet};
    \node [dashblock, below of=dev] (amt) {AMT Crowd Workers};
    \path [line] (init) -- (sim);
    \path [line] (sim) -- (init);
    \path [line] (init) -- (dev);
    \path [line] (dev) -- (init);

\end{tikzpicture}
    \caption{A depiction of our two audit methods: crowdsourced auditing via Amazon Mechanical Turk workers, and sock puppet auditing via an Appium-controlled iPhone simulator.}
    \label{fig:my_label}
\end{figure}

\subsection{Mechanism Audit}

\subsubsection{Experiment 1a: Measuring platform-wide update frequency.}\label{platform-wide}
We identify two content update mechanisms on Apple News: platform-wide updates and user-specific updates. In platform-wide updates, the ``master list'' of Trending Stories changes on the Apple News servers. The two types of updates may not always correlate: an update to the master list may not cause an update to the stories seen by a given user. We tested the platform-wide update frequency by continually checking for new content and measuring how often content changed. In this experiment, we set Apple News to not use any personalization data from other apps by turning off the ``Find Content in Other Apps'' toggle in the settings. We also removed files from the app's cache folder before every refresh. These measures were intended to minimize any potential content adaptation while we investigated the platform-wide update frequency of the Trending Stories section. However, the measures took time to perform, as did the process of opening the app, locating the buttons, and pressing the buttons via Appium. Thus, the maximum frequency with which we could record Trending Stories this way was approximately every 2 minutes.

\subsubsection{Experiment 1b: Measuring user-specific update frequency.}\label{user-specific}
Our initial experiment to test for user-specific update frequency happened by accident when we collected data for 12 continuous hours without closing the app and observed no changes in the Trending Stories. This led us to ask: Under what conditions does Apple News allow new Trending Stories from the ``master list'' to populate a user's app? In other words, what might be the user-specific update frequency? To answer this, we performed the same process as in experiment 1a, continuously collecting data from the app, but \textit{without deleting the cache and user identification files}. This allowed us to determine whether Apple News updates Trending Stories on a different schedule for individual users compared to the ``master list.''

\subsubsection{Experiment 2: Testing for location-based adaptation.}
While Apple News is known to adapt stories at a national level \citep{Brown2018}, we wondered whether stories changed with respect to a more fine-grain measure of location such as city or state. To test for fine-grain localization, we designed an experiment to elicit differences in content that could be accounted for by differences in a device's location.

In this experiment, we gathered stories from different simulated locations via Appium. Because it was not possible to run 50 simulators in parallel, we designed a way to sequentially collect and analyze data for evidence of localization, while still controlling for differences owed to time. We ran two iPhone simulators in parallel: one control, and one experimental. The control simulator was virtually located at Apple's headquarters and collected a set of control headlines, while the experimental simulator was virtually moved around the country to the 50 state capitals to collect local headlines in each place. Both simulators collected all Trending Stories shown in the app. We defined location-based adaptation (i.e. localization) as local headlines differing from  control headlines at a single point in time. Algorithm \ref{localization_algorithm} details the experiment.

\begin{algorithm}
\caption{Check for localization via sock-puppeting}\label{localization_algorithm}
\begin{algorithmic} 
\REQUIRE{set of locations $L$, $control\_location$}
\STATE set control simulator location to $control\_location$
\FOR{each $location$ in $L$}
\STATE set experimental simulator location to $location$
\STATE collect $local\_headlines$ and $control\_headlines$
\IF{$control\_headlines$ != $local\_headlines$}
\RETURN True \COMMENT{control headlines differed from local headlines, thus localization}
\ENDIF
\ENDFOR
\RETURN False \COMMENT{No localization observed}
\end{algorithmic}
\end{algorithm}

\subsubsection{Experiment 3: Measuring user-based adaptation.}
Location is one variable that might be used to adapt content across users. However, in Apple News, individual users can ``follow'' specific topics and publishers. This feature is used to populate the app with personalized content outside of the Trending Stories and Top Stories sections, but it is unclear whether it may also be used to adapt stories in the Trending Stories section. Because of the social and political implications of personalizing content, we wanted to test whether any personalization was applied to the algorithmically-curated Trending Stories section -- we do not examine Top Stories for personalization because Apple's editorial team is known to ``select five stories to lead the app'' in the United States \citep{Nicas2018}. In the personalization experiment, we used AMT to collect synchronized user screenshots, and compared the list of headlines to a control list that we collected concurrently from the simulator (which contained no cache files or user profile data).

Meaningful quantitative comparison of ranked lists is nontrivial. As \cite{Webber2010} point out, ``testing for statistical significance becomes problematic,'' since ``any finding of difference'' disproves a null hypothesis that rankings are identical. Researchers have therefore proposed and used various metrics to quantify the \textit{degree} of similarity between ranked lists, depending on the characteristics of the list. For example, rank-biased overlap (RBO) is designed for lists in which (1) the length is indefinite and (2) items at the top of the list are more important than items in the tail \citep{Webber2010}, and can be useful, for example, to investigate the degree of personalization in search engines  \citep{Robertson2018a}. 

While the Top Stories and Trending Stories lists vary in length across devices, their lengths are not ``indefinite'' since there is a known maximum length. Further, the weighting schemes in metrics such as RBO do not conceptually fit our data. This is because, as previously mentioned, one subset of Trending Stories (\#1-2) is displayed on iOS home screens, one subset (\#1-4) is shown within the app on all devices, and the full set of Trending Stories (\#1-6) is shown within the app on devices with large screens. Thus, a headline moving \textit{between} these subsets (ex. from \#5 to \#4) has more significant implications because the change will more substantively affect visibility compared to a headline moving \textit{within} a subset (e.g., from \#4 to \#3).

We therefore define user-based adaptation as a unique \textit{set} of stories appearing to a given user, quantified by the overlap coefficient (Szymkiewicz-Simpson coefficient)  \citep{Vijaymeena2016} between a user's Trending Stories and the Trending Stories on a control device. The overlap coefficient is a slight modification of Jaccard index, which is also used by \citet{Hannak}, \citet{Kliman-Silver2015}, \citet{Vincent2019}, and other studies to measure personalization in web search results. It operationalizes the definition of user-based adaptation as the size of the intersection divided by the smallest size of the two sets, and ranges from 0 (no overlap between sets) to 1 (one set includes all items in the other set). One key benefit of the metric is that it conceptually accommodates data from devices with smaller screens that showed only four Trending Stories, since our control list was on a large device and showed six Trending Stories.

\subsection{Content Audit}

\subsubsection{Experiment 4: Extended data collection.}
Finally, we designed and performed an extended data collection using Appium based on findings from the first three experiments. Each data point included a timestamp and a list of shortened URLs to the news stories, gathered by a simulated user tapping ``Share'' then ``Copy'' on each headline. To analyze the stories, a Python script followed each shortened URL along redirects until it reached a final web address, where it parsed the web page for a story title. The domain of the final address identified the publisher of the story (e.g. \texttt{cnn.com}).


We compared the Top Stories and Trending Stories according to the update schedule, source diversity, and headline content. To quantify source diversity, we relied on the Shannon diversity index \citep{Shannon1948}, a metric which accounts for both abundance and evenness of different categories in a sample and derives from the probability that two randomly-chosen items belong to the same category (in our case, the same news source). Normalizing this value (dividing it by the maximum possible Shannon diversity index for the population) provides the Shannon \textit{equitability} index, also known as Pielou's index \citep{Pielou1966}, a more interpretable value between 0 and 1 where 1 indicates that all sources appeared with the same relative frequency (e.g., 50\% of Trending Stories from Fox and 50\% of Trending Stories from CNN). To statistically compare the source diversity of the Trending Stories section with that of the Top Stories section, we use the Hutcheson t-test \citep{Hutcheson1970}, which was specifically designed to compare the Shannon diversity index between two samples (in our case, the Trending Stories and the Top Stories).


To compare content between the two sections, we first create a corpus of all headlines from Top Stories, and another corpus of all headlines from Trending Stories. We then count bigrams and trigrams within each set of headlines. Before doing this, each headline is stripped of punctuation and stopwords, then tokenized into bigrams and trigrams. We calculate the log ratio \citep{Hardie2014} of each n--gram to determine which are most salient relative to the other section.


\section{Results}

\subsection{Mechanism}
\subsubsection{Experiment 1: Update Frequency of Trending Stories}
To measure the system-wide update frequency, we ran the automated data collection every two minutes over a 24-hour period, closing the app and erasing the cache folder each time the simulator collected stories. The average time between updates -- either a change in the order of stories or the addition of a new story -- was 20 minutes (min=6 mins, max=61 mins, median=17 mins, sd=11 mins). We ran the user-specific update frequency experiment during the same time (closing the app but not removing cache and user profile files) and found identical update times. However, when the simulator kept the app open in between checking for new stories, the Trending Stories section did not change for the full 24 hours.

\subsubsection{Experiment 2: Localization of Trending Stories}
We ran our experiment to test for localization on two occasions, finding no variation in trending stories that could be attributed to differences in location. In other words, at all 50 experimental locations, the list of Trending Stories in the experimental location matched the list of Trending Stories in the control location, thus showing no evidence of location-based content adaptation. The experiment took about an hour and a half to run.

\subsubsection{Experiment 3: Personalization of Trending Stories}
We collected synchronized screenshots on three occasions, yielding 28 data points in the first experiment, 32 in the second, and 29 in the third for a total of 89 screenshots from real-world users. 83 screenshots (93.3\%) displayed the same time as the control screenshot (adjusted for time zone), five screenshots (5.6\%) displayed times one minute later, and one screenshot (1.1\%) displayed a time three minutes later. We exclude these six mistimed screenshots in our analysis.

We compared headlines in user screenshots to control headlines collected from the non-personalized iPhone simulator at the same time. Across the 83 screenshots examined, the average overlap coefficient between the experimental headlines and the control headlines was 0.97, indicating that the vast majority of screenshots contained the same headlines as the control. Examining the data more closely, 74 user screenshots (89.2\%) showed headlines that all appeared in the control headlines, while 9 screenshots (10.8\%) contained a single headline that did not appear in the control headlines at the same time.

Since 9 user screenshots included a headline that did not appear in the control headlines at the same time, we examined whether the differences correlated to the user-provided zip code, user time zone, the device's carrier, or the device screen size, but we observed no pattern. We then expanded the Trending Stories from our control device to include headlines that appeared in the preceding set of control headlines (i.e. the control set of Trending Stories before its most recent change). When we included these headlines, the average overlap coefficient between the experimental headlines and the control headlines was 1.00. In other words, no headlines were unique to a given user, because all headlines in the screenshots also appeared in the control headlines.

These experimental results suggest Apple News does not implicitly personalize trending stories: in the 10.8\% of screenshots that did not fully match the control headlines, just one headline was different, and including Trending Stories from the most recent prior set of control headlines yielded \textit{no unique headlines across all 83 screenshots}.

We note that despite this lack of personalization, Apple News users may still see different headlines in Trending Stories for various reasons. First, as exemplified above, Trending Stories appear to refresh with minor delays in some cases. Secondly, we found that if a source was blocked in an experimental simulator, Trending Stories would not include stories from that source even when the Trending Stories in the control simulator did include such stories. Finally, devices with smaller screens (iPhone 6, 7, SE, X, and XS) show a list of four Trending Stories, while devices with larger screens (iPhone XS Max, all iPads, and all MacOS devices) show a list of six Trending Stories.


\subsection{Content}

To examine prominent content in Apple News and compare the app's human curation and algorithmic curation, we ran extended data collection of Trending Stories and Top Stories beginning at 12:01am on March 9th and ending at 11:59pm on May 9th 2019 (62 days), collecting a total of 1,268 Top Stories and 3,144 Trending Stories. In this section, we present and compare the results from each section.

\subsubsection{Source Concentration in Trending Stories vs. Top Stories}

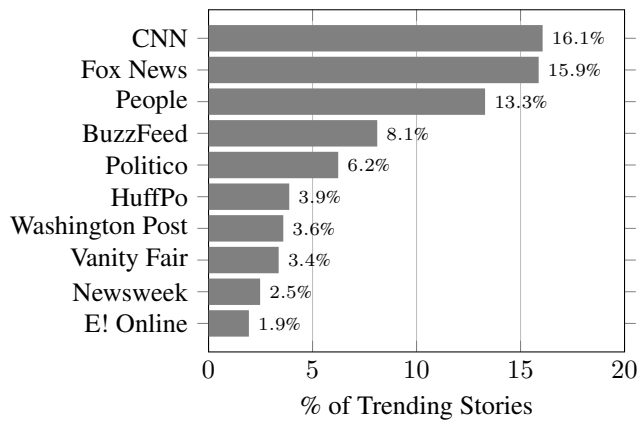
\begin{figure}[t]
\begin{center}
\begin{tikzpicture}
\begin{axis}[ 
width=0.4\textwidth,
xbar, xmin=0, xmax=20,
xmajorgrids=true,
nodes near coords=\pgfmathprintnumber{\pgfplotspointmeta}\%,
every node near coord/.append style={font=\scriptsize},
nodes near coords style={/pgf/number format/.cd,fixed zerofill,precision=1},
xlabel={\% of Trending Stories},
symbolic y coords={%
{E! Online},
{Newsweek},
{Vanity Fair},
{Washington Post},
{HuffPo},
{Politico},
{BuzzFeed},
{People},
{Fox News},
{CNN} },
ytick=data,
]
\addplot [fill=gray,draw=none]
coordinates {
(16.062341,{CNN})
(15.8715013,{Fox News})
(13.2951654,{People})
(8.110687,{BuzzFeed})
(6.2340967,{Politico})
(3.8804071,{HuffPo})
(3.5941476,{Washington Post})
(3.3715013,{Vanity Fair})
(2.480916,{Newsweek})
(1.9402036,{E! Online})
};
\end{axis}
\end{tikzpicture}
\end{center}
\caption{Relative distribution of Trending Stories across top ten sources, March 9th to May 9th, 2019 (n=3,144)}\label{fig-trending_stories_distribution}
\end{figure}

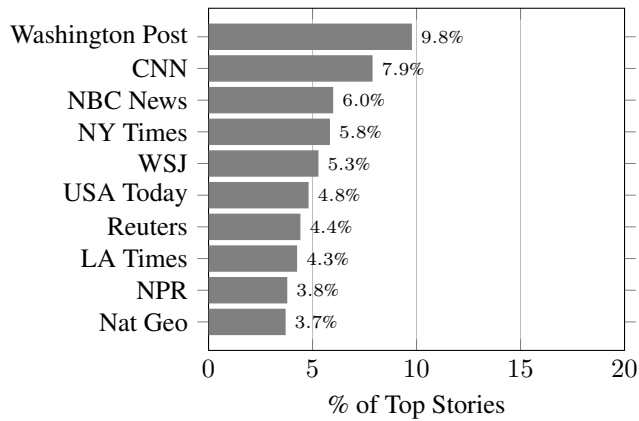
\begin{figure}[!t]
\begin{center}
\begin{tikzpicture}
\begin{axis}[ 
width=0.4\textwidth,
xbar, xmin=0, xmax=20,
xlabel={\% of Top Stories},
xmajorgrids=true,
nodes near coords=\pgfmathprintnumber{\pgfplotspointmeta}\%,
every node near coord/.append style={font=\scriptsize},
nodes near coords style={/pgf/number format/.cd,fixed zerofill,precision=1},
symbolic y coords={%
{Nat Geo},
{NPR},
{LA Times},
{Reuters},
{USA Today},
{WSJ},
{NY Times},
{NBC News},
{CNN},
{Washington Post}
},
ytick=data,
]
\addplot [fill=gray,draw=none]
coordinates {
(9.7791798,{Washington Post})
(7.8864353,{CNN})
(5.9936909,{NBC News})
(5.8359621,{NY Times})
(5.2839117,{WSJ})
(4.8107256,{USA Today})
(4.4164038,{Reuters})
(4.2586751,{LA Times})
(3.785489,{NPR})
(3.7066246,{Nat Geo})
};
\end{axis}
\end{tikzpicture}
\end{center}
\caption{Relative distribution of Top Stories across top ten sources, March 9th to May 9th, 2019 (n=1,268)}\label{fig-top_stories_distribution}
\end{figure}

A key finding from our content analysis is the high concentration of sources in the algorithmically-curated Trending Stories section. Of the 83 unique sources observed in Trending Stories, CNN was the most common, with 505 unique stories (16.1\%) in the collection. The distribution across sources of Trending Stories was also highly skewed, as the three most common sources accounted for 1,422 stories (45.2\%). In contrast, out of 87 unique sources observed in Top Stories, the most common source accounted for 124 unique stories (9.8\%), and the three most common sources accounted for 300 stories (23.7\%). Across sections 40 sources appeared in both. While the Trending Stories achieved a Shannon equitability index of 0.689, the Top Stories achieved an index of 0.780. The Shannon indices indicate a significant difference (Hutcheson's t = 11.17, p < 0.001) in the diversity and evenness of sources between Top Stories and Trending Stories, as can be visualized in Figures \ref{fig-trending_stories_distribution} and \ref{fig-top_stories_distribution}. See Table \ref{summary-table} for a further comparison of source prominence and source diversity.

\pgfplotstableread[row sep=\\,col sep=&]{
hour & pct   \\
0     & 4.2302799   \\
1     & 3.8804071   \\
2    & 3.5941476   \\
3   & 3.148855   \\
4   & 3.0534351   \\
5   & 3.8486005   \\
6   & 4.1666667   \\
7     & 4.7391858   \\
8     &  5.0572519   \\
9    & 5.8206107   \\
10   &  5.8206107   \\
11   &  4.1666667   \\
12   &  4.1030534   \\
13     &  3.6259542   \\
14    &  4.3256997   \\
15   & 3.8486005   \\
16   &  3.6895674   \\
17     &  4.1030534   \\
18     &  3.9122137   \\
19    &  3.7849873   \\
20   &  4.5483461   \\
21   &  3.8167939   \\
22   &  4.0076336   \\
23   &  4.7073791   \\
}\myhrdata

\begin{figure}[t]
\begin{center}
\begin{tikzpicture}
    \begin{axis}[
            ybar,
            ymajorgrids=true,
            enlargelimits=false,
            ymin=0,ymax=15,
            yticklabel=$\pgfmathprintnumber{\tick}$\%,
            bar width=0.5,
            xlabel={Hour of Day (PDT)},
            enlarge y limits={0.0,upper},
            enlarge x limits={0.04},
            height=0.2\textwidth,
            width=0.5\textwidth,
            xtick distance=4,
            ]
        \addplot[fill=gray,draw=none] table[x=hour,y=pct]{\myhrdata};
    \end{axis}
\end{tikzpicture}
\end{center}
\caption{Relative distribution of Trending Stories over the hours when they first appeared. New stories appear consistently throughout the day, peaking slightly at around 10am.}\label{fig-trending-stories-appearance-times}
\end{figure}
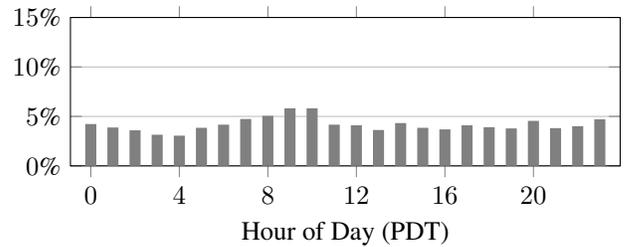

\pgfplotstableread[row sep=\\,col sep=&]{
hour & pct   \\
0     & 4.2586751 \\
1     & 1.0252366   \\
2    & 0.7097792   \\
3   & 0.4731861   \\
4   & 0   \\
5   & 0.4731861   \\
6   & 1.340694   \\
7     & 7.2555205   \\
8     &  5.4416404   \\
9    & 4.1798107   \\
10   &  5.7570978   \\
11   &  2.681388   \\
12   &  1.1829653   \\
13     &  2.8391167   \\
14    &  13.4069401   \\
15   & 5.4416404   \\
16   &  3.5488959   \\
17     &  1.4195584   \\
18     &  2.9968454   \\
19    &  13.6435331   \\
20   &  6.1514196   \\
21   &  2.5236593   \\
22   &  2.0504732   \\
23   &  11.1987382   \\
}\myhrdata
    
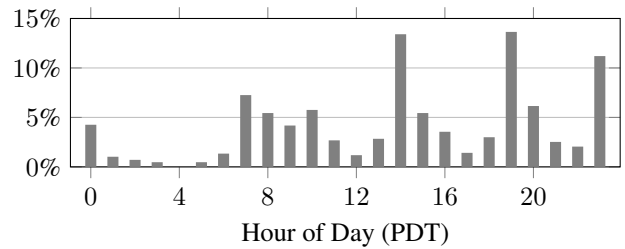
\begin{figure}[t]
\begin{center}
\begin{tikzpicture}
    \begin{axis}[
            ybar,
            ymajorgrids=true,
            bar width=0.5,
            enlarge y limits={0.0,upper},
            enlarge x limits={0.04},
            ymin=0,ymax=15,
            xlabel={Hour of Day (PDT)},
            yticklabel=$\pgfmathprintnumber{\tick}$\%,
            height=0.2\textwidth,
            width=0.5\textwidth,
            xtick distance=4,
        ]
        \addplot[fill=gray,draw=none] table[x=hour,y=pct]{\myhrdata};
    \end{axis}
\end{tikzpicture}
\end{center}
\caption{Relative distribution of Top Stories over the hours when they first appeared. New stories appear more commonly at specific times in the day, such as 2pm or 7pm. }\label{fig-top-stories-appearance-times}
\end{figure}

\subsubsection{Update Patterns in Trending Stories vs. Top Stories}
Data from the extended collection suggests that Trending Stories follow a fairly consistent churn rate, with stories gradually trickling in throughout the day, while the Top Stories section receives punctuated updates in the morning, mid-day, afternoon, and evening. These update schedules are visualized in Figures \ref{fig-trending-stories-appearance-times} and \ref{fig-top-stories-appearance-times}.

\subsubsection{Topics in Trending Stories vs. Top Stories}
Our n--gram analysis highlights topical distinctions between the app's editorially-curated content and algorithmically-curated content. For example, the Trending Stories (see Table \ref{trending-only-ngrams}) frequently featured many celebrities (Lori Loughin, Kate Middleton, Justin Bieber, Olivia Jade, Stephen Colbert, Kim Kardashian), as well as some politicians (Donald Trump, Alexandria Ocasio-Cortez). Ocasio-Cortez never appeared in Top Stories, suggesting a mismatch between her trending popularity and the interests of editors; however, Trump appeared in both lists, indicating popularity-driven interest as evidenced by his full name trending, and also editorial interest in his activities as president of the United States. The Trending Stories section tended to feature more news that might be considered surprising, shocking, or sensational, with frequent terms such as ``found dead'' and ``florida man'' (referencing an Internet meme about Florida's purported notoriety for strange and unusual events\footnote{https://en.wikipedia.org/wiki/Florida\_Man}). On the other hand, the salient terms found in Top Stories (see Table \ref{top-only-ngrams}) featured more substantive policy issues related to topics like health care (e.g. ``affordable care act''), immigration (e.g. ``border wall'' and ``sanctuary city''), and international politicians and events (e.g. ``Kim Jong Un'' and ``brexit deal'') which never appeared in the Trending Stories headlines. Also, salient n--grams from Top Stories reflect news roundups which the editors feature regularly (e.g. ''biggest news'' and ``week's good news''), as well as some entertainment news (e.g. ''Avengers: Endgame'').

\begin{table}[!t]
\small
    \begin{subtable}{0.25 \textwidth}
      \centering
        \caption{Trending Stories}
        \begin{tabular}{@{}ll@{}}
            \toprule
            N--gram    & n \\
            \midrule
donald trump     & 65     \\
fox news     & 29     \\
donald trump's     & 25     \\
lori loughlin's     & 18     \\
kate middleton     & 16     \\
justin bieber     & 14    \\
found dead     & 13     \\
olivia jade     & 13      \\
alexandria ocasio-cortez     & 12     \\
prince william     & 12       \\
trump jr    & 12      \\
meghan markle     & 33*      \\
florida man     & 11     \\
ivanka trump     & 11      \\
jared kushner     & 11       \\
kim kardashian     & 11       \\
south carolina     & 11      \\
stephen colbert     & 11      \\
things that'll     & 11     \\
tucker carlson     & 11     \\

            \bottomrule
            \end{tabular}
    \label{trending-only-ngrams}
    \end{subtable}%
    \begin{subtable}{0.25 \textwidth}
      \centering
        \caption{Top Stories}
        \begin{tabular}{@{}ll@{}}
            \toprule
            N--gram    & n \\
            \midrule
    biggest news        & 9    \\
    measles cases        & 9    \\
    `sanctuary cities'        & 8    \\
    attorney general barr        & 8    \\
    affordable care act       & 7    \\
    trump threatens      & 7     \\
    2020 presidential      & 6   \\
    avengers: endgame       & 6    \\
    house panel       & 6    \\
    new zealand mosque       & 6    \\
    notredame fire       & 6    \\
    trump tax returns       & 6    \\
    week's good news       & 6    \\
    attorney general        & 16*    \\
    border wall        & 5    \\
    brexit deal        & 5    \\
    ground boeing 737        & 5    \\
    julian assange        & 5    \\
    kim jong un        & 5    \\
    timmothy pitzen        & 5    \\
            \bottomrule
            \end{tabular}
        \label{top-only-ngrams}
    \end{subtable} 
        \caption{The twenty most-salient n--grams in Trending Stories and Top Stories (as measured by the log ratio of occurences in each section's headlines), and the raw number of occurrences. In almost all cases, n--grams were unique to the section. Otherwise, n* indicates the n--gram appeared twice in the other section.}
\end{table}

\begin{table}[h]
\resizebox{0.47 \textwidth}{!}{
\begin{tabular}{@{}lll@{}}
\toprule
                    & Trending Stories & Top Stories \\ \midrule
Curation            & Algorithmic                 & Editorial          \\
Stories Displayed &  6 (4 on small screens)   &  5 \\
Localization        & National                    & National               \\
Personalization     & No                        &     No                  \\
Total Stories Analyzed & 3,144 & 1,267 \\

\midrule
Avg. Story Duration &  2.9hrs & 7.2hrs  \\

Avg. Stories per Day  &   50.7  &  20.4     \\   

Avg. Stories per Day per Slot  &   8.5  &  4.1   \\   
\midrule

Total Unique Sources & 83 & 87 \\

Shannon Equitability Index & 0.689 & 0.780 \\

Mean Source Share  &   1.2\%  &  1.1\%     \\   

Median Source Share  &   0.2\%  &  0.2\%     \\ 

Top Source Share      &    16.1\%      & 9.8\%          \\   

Top 3 Sources Share      &    45.2\%      & 23.7\%          \\   

Top 10 Sources Share      &    74.8\%      & 55.7\%          \\  
\bottomrule
\end{tabular}%

}
\caption{Summary of results. The top section focuses on high-level aspects, the middle on churn rate, and the lower on source distribution. ``Top N Source Share'' is defined as the percentage of stories that came from the N most common sources in the section.} \label{summary-table}
\end{table}

\section{Discussion}\label{sec:discussion}
In this section we first discuss the specific results found in our audit of Apple News and how our results paint a distinction between algorithmic logic and human editorial logic. We then consider the broader implications of our study in terms of the various audit methods used, clarifying their different strengths. And finally, we suggest that the audit framework we develop may contribute to informing a conceptual ``audit standard'' which helps make the questions and methods more consistent when approaching similar audits of other news curators.

\subsection{Mechanisms behind Apple News}
The absence of localization and personalization in the sections we audited highlights a possible tension between equitable economic distribution and vulnerability to echo-chambers. By showing the same Top Stories and Trending Stories to every user, Apple has taken a measure that minimizes individual filter bubble effects. However, this design choice may also translate to a highly-skewed distribution of economic winnings for publishers, since the few sources that frequent the top slots reap disproportionate traffic and potential advertising revenue from the 85 million users. 

Certainly, a balance could mitigate filter bubble effects while creating more equitable monetization. For example, Google News provides all users with an algorithmically-selected five-story briefing, but the same two top stories are shown to all users \citep{Wang2018}. Apple might explore a similar hybrid approach, using their editorial team to choose a mix of regionally and nationally relevant news. It should also be noted that Apple News includes a ``For You" section based on ``topics \& channels you read," which may help balance these tensions.

\subsection{Algorithmic vs. Editorial Logic}
Our results illustrate what \citet{Gillespie2014} refers to as a possible competition between ``editorial logic'' and ``algorithmic logic.'' While the algorithmic logic behind Trending Stories is intended to ``automate some proxy of human judgment or unearth patterns across collected social traces,'' the editorial logic behind Top Stories ``depends on the subjective choices of experts, themselves made and authorized through institutional processes of training and certification'' \citep{Gillespie2014}. These logics manifest in our audit results for both mechanism and  content: Trending Stories and Top Stories exhibit distinct update schedules, sources, and topics that reflect their respective logics.

According to interviews with Apple's staff by \citet{Nicas2018}, the editorial logic behind the Top Stories pushes new content ``depending on the news,'' and also ``prioritizes accuracy over speed.'' In contrast, the Trending Stories section continually churns out new content (over 50 stories per day on average, more than twice as many as the Top Stories section). The algorithmic logic continually updates `what's trending,' and does so at all hours of the day (see Figure \ref{fig-trending-stories-appearance-times}), whereas the editorial logic espoused by Apple's staff strives for ``subtly following the news cycle and what's important'' \citep{Nicas2018}.

The contrast in logics extends to source concentration and source diversity. The editorially-selected Top Stories section exhibited more diverse and more equitable source distribution than the algorithmically-selected Trending Stories, as detailed in Table \ref{summary-table} and visualized in Figures \ref{fig-trending_stories_distribution} and \ref{fig-top_stories_distribution}. During the two-month data collection, human editors chose from a slightly wider array of sources than the algorithm behind Trending Stories, and also distributed selections more equitably across those sources, although there was a core set of 40 sources that were selected by \textit{both} algorithm and editor. Apple's editors also appeared to seek out regional news sources when the topic called for it. For example, during our data collection, the team chose stories from The San Diego Union-Tribune, The Miami Herald, The Chicago Tribune, TwinCities.com, and The Baltimore Sun. While the Trending Stories algorithm surfaced some content from smaller sources (e.g., esimoney.com, a single-author website dedicated to money management), no sources were regionally-specific.

Finally, perhaps the most intuitive distinction between the editorial logic and the algorithmic logic exhibited in Apple News is differences in content. Trending Stories uniquely included ``soft news'' \citep{Reinemann2012} pieces about celebrities and entertainment (ex. Kate Middleton, Justin Bieber), while the Top Stories uniquely included ``hard news'' topics including international stories and news about political policy (ex. Brexit, Affordable Care Act). Our data corroborates initial reports that headlines in Trending Stories ``tend to focus on Mr. Trump or celebrities'' \citep{Nicas2018}.

\subsection{Broader Implications}

\subsubsection{Audit Methods}
To study Apple News we used scraping, sock-puppet, and crowdsourced auditing. While these techniques have been employed in previous audit studies, we next reflect on our experience of how we found each technique helpful for addressing different aspects of our audit framework, in hopes this can inform future audit studies. 

First, audit studies should consider crowdsourcing whenever real-world observation is important, or when seeking higher parallel throughput in the data collection process. By using the crowd, we were able to assess the degree of personalization Apple News performs in practice, rather than fully relying on simulated sock-puppet data. Also, since resource constraints limited us to run at most two simulators at a time, crowdsourcing allowed us to significantly increase the throughput of our data collection, as many crowd workers could take screenshots in parallel.

Sock-puppet auditing -- using computer programs to impersonate users \citep{Sandvig2014} -- is most helpful for isolating variables that might affect a given system. Sock puppets provide clear and precise information about the input to an algorithm in cases where the crowd falls short. For example, precise temporal synchronization was still challenging with crowdworkers, as 6.7\% of screenshots from the crowd showed times that did not match the requested time in minutes, and even screenshots at the correct hour and minute may have been unsynchronized if seconds were taken into account. On the other hand, when using sock puppets, we could perform time-locked data collection on multiple devices with same-second precision.

Lastly, we found the scraping audit most helpful for extended data collection, and we suggest scraping whenever researchers seek continuous data or to monitor over time. While we initially deployed a crowdsourced method for the extended data collection \citep{Bandy}, we found that we could not rely on the crowd for consistent data over time. Namely, in the United States, it was difficult to collect screenshots between the hours of 1am and 5am. However, after observing no evidence of personalization or localization in our initial experiments, we needed data from just one user account. We could therefore scrape content from a single simulated device to run the extended data collection.

It should be noted that we categorize our Appium-based data collection as a scraping audit since it centers around ``repeated queries to a platform and observing the results'' \citep{Sandvig2014}, however, we used a simulated iPhone to impersonate a user, making it somewhat of a hybrid with sock-puppet auditing. The Apple News platform lacks a public or private endpoint from which to scrape stories, so our experiment required additional layers of operation -- every data point we collected required simulating a user opening the application, refreshing for new content, locating buttons, and pressing buttons, thus requiring more time to collect data compared to a traditional scrape.

\subsubsection{Audit Framework}
To guide this work, we developed a conceptual framework that articulates three common aspects of a curation system that an audit might address: mechanism, content, and consumption. We showed how each of these aspects can have consequences for individual users, publishing companies, and even political discourse. As news curation systems change, consequential aspects may also change and prompt an expanded or revised framework.

Still, future research might leverage our proposed framework for auditing other content curators, revising and elaborating it to suit the nuances of specific systems. We believe this framework helps advance towards a conceptual ``audit standard'' for curation systems, which might allow the research community to compare and contrast different curation platforms, as well as characterize the evolution of a single platform over time.

\section{Conclusion}
This paper introduced, motivated, and applied an audit framework for news curation systems. We showed how a system's mechanism and content have personal, political, and economic implications in society, and developed a specific research framework accordingly. To demonstrate this framework, we focused on Apple News. To our knowledge, we are the first to analyze this platform in-depth. We employed several audit methods to analyze the app's update frequency, content adaptation, source concentration, and more. Results showed that the human-curated Top Stories section features fewer stories per day and exhibits greater source diversity and greater source evenness compared to the algorithmically-curated Trending Stories section. We discussed how these differences in mechanism and content reflect differences between the algorithmic and editorial curation logics underpinning the two sections.  We offer our framework for reuse, revision, and elaboration to any researchers studying sociotechnical news curation systems. 

\section{Acknowledgements}
This work is funded in part by the National Science Foundation under award number IIS-1717330.

\small{
\bibliography{library}
\bibliographystyle{aaai}
}
\end{document}